\documentclass[proof]{pasj00}

\begin{document}
\SetRunningHead{Author(s) in page-head}{Running Head}
\Received{2000/12/31}
\Accepted{2001/01/01}

\title{A Peculiar Jet and Arc of Molecular
Gas Toward the Rich and Young Stellar Cluster
Westerlund 2 and a TeV Gamma Ray Source}

\author{Yasuo \textsc{Fukui},\altaffilmark{1}
Naoko \textsc{Furukawa},\altaffilmark{1}
Thomas M. \textsc{Dame},\altaffilmark{2}
Joanne R. \textsc{Dawson},\altaffilmark{1,3}
Hiroaki \textsc{Yamamoto},\altaffilmark{1}
Gavin P. \textsc{Rowell}, \altaffilmark{4}
Felix \textsc{Aharonian},\altaffilmark{5,6}
Werner \textsc{Hofmann},\altaffilmark{6}
Emma de O$\tilde{\rm n}a$ \textsc{Wilhelmi},\altaffilmark{6}
Tetsuhiro \textsc{Minamidani},\altaffilmark{7}
Akiko \textsc{Kawamura},\altaffilmark{1}
Norikazu \textsc{Mizuno},\altaffilmark{1, 8}
Toshikazu \textsc{Onishi},\altaffilmark{1}
Akira \textsc{Mizuno},\altaffilmark{9}
Shigehiro \textsc{Nagataki} \altaffilmark{10}  %
  }

\altaffiltext{1}{Department of Astrophysics, Nagoya University, Furocho, Chikusa-ku, Nagoya, Aichi, 464-8602}
\altaffiltext{2}{Harvard-Smithsonian Center for Astrophysics, 60 Garden Street, Cambridge MA 02138, USA.}
\altaffiltext{3}{Australia Telescope National Facility, CSIRO, P.O. Box 76, Epping NSW 1710, Australia}

\altaffiltext{4}{School of  Chemistry and Physics, University of Adekaide, Adelaide 5005, Australia.}
\altaffiltext{5}{Dublin Institute for Advanced Studies, 31 Fitzwilliam Place, Dublin 2, Ireland.}
\altaffiltext{6}{Max-Planck-Institut fur Kernphysik, PO Box 103980, 69029 Heidelberg, Germany.}
\altaffiltext{7}{Department of Physics, Faculty of Science, Hokkaido University, N10W8, Kita-ku, Sapporo 060-0810}
\altaffiltext{8}{National Astronomical Observatory of Japan, Osawa, Mitaka, Tokyo 181-8588, Japan.}
\altaffiltext{9}{Solar-Terrestrial Environment Laboratory, Nagoya University, Furocho, Chikusa-ku, Nagoya 464-8601}
\altaffiltext{10}{Yukawa Institute for Theoretical Physics, Kyoto University, Sakyo-ku, Kyoto 606-8502}

\email{fukui@a.phys.nagoya-u.ac.jp}


%


\maketitle

\begin{abstract}
We have discovered remarkable jet- and arc-like molecular features
toward the rich and young stellar cluster Westerlund2. The jet has
a length of $\sim100$ pc and a width of $\sim10$ pc, while the arc
shows a crescent shape with  a radius of $\sim30$ pc. These
molecular features each have masses of $\sim10^4 M_{\solar}$ and
show spatial correlations with the surrounding lower density HI gas.
The jet also shows an intriguing positional alignment with the core
of the TeV gamma ray source HESS J1023-575 and with the MeV/GeV
gamma-ray source recently reported by the $Fermi$ collaboration.
We argue that the jet and arc are  caused  by  an energetic event
in Westerlund 2, presumably due to  an anisotropic supernova
explosion of one of the most massive member stars.  While the
origin of the TeV and GeV gamma-ray sources is uncertain, one may
speculate that they are related to the same event via relativistic
particle acceleration by strong shock waves produced at the explosion
or by remnant objects such as a pulsar wind nebula or microquasar.
\end{abstract}

\section{Introduction}
Massive stars have a strong influence on the surrounding 
interstellar space via their stellar winds and ultraviolet 
radiation.  Moreover, they end their lives in catastrophic 
supernova (SN) explosions, thus providing more energetic 
impact on the interstellar medium.
Wd2 is a rich star cluster of age $2 - 3$ Myrs \citep{pia98}. 
CO observations have allowed a distance estimate to the 
source of $5.5\pm1.5$ kpc (\cite{fur09}; see also \cite{dam07}). 
The total stellar mass in Wd2 is of the order of 
$4500 M_{\solar}$, including 12 O stars and 2 WR stars 
\citep{rau07}. It is also associated with a HII region 
RCW49, a remarkable infrared nebula as revealed by 
$Spitzer$ \citep{chu04}, and perhaps with the extended 
TeV gamma-ray source HESS J1023-575, discovered with the 
HESS telescope array in the direction of the cluster 
\citep{aha07}.  Recently, low energy (MeV/GeV) gamma-rays 
have been reported from the same direction by the $Fermi$ 
collaboration \citep{abd09}.

In this Letter we describe the discovery of a spectacular jet and arc of
molecular gas detected with the NANTEN telescope in the $^{12}$CO($J=1-0$)
mm-wave emission line survey \citep{miz04}, and discuss their possible links 
to the MeV/GeV and TeV gamma-ray sources.

\section{Results}
We show the two new CO features toward Wd2 in Figure 1, 
where the CO distribution in the velocity range $\sim20-30$ 
km s$^{-1}$ is presented in Galactic coordinates. One is a 
straight feature in the east at $(l, b) = 
(284^{\circ}.2 - 284^{\circ}.9, -1^{\circ}.0 - -0^{\circ}.4)$ 
(hereafter, the jet) and the other an arc-like feature in the 
west at $(l, b) = (283^{\circ}.9 - 284^{\circ}.3, -0^{\circ}.5 - 0^{\circ}.0)$ 
(hereafter, the arc). We note that the arc has an intensity peak and a protrusion 
to the west of HESS J1023-575 at $(l, b) = 
(284^{\circ}.04 - 284^{\circ}.09, -0^{\circ}.31 - -0^{\circ}.24)$, 
both of which lie along the jet axis, suggesting the existence of 
the diametrically opposite part of the jet.

These two features show a remarkable correlation with Wd2 / HESS J1023-575. 
The jet is extended over $1^{\circ}$ to the east of Wd2 / HESS J1023-575, 
tilted at an angle of $\sim39^{\circ}$ to the galactic plane, and is very 
well aligned with Wd2 / HESS J1023-575 as well as a ridge of the radio 
continuum corresponding to the HII region, RCW 49\citep{whi97}. 
A least-squares fitting to the jet in Figure 1 yields a regression,

\begin{equation}
b + 1^{\circ}.0  = (\timeform{-2'.50} \pm \timeform{0'.20}) + (0.81 \pm 0.01) (l - 285^{\circ}.0),
\end{equation}

\noindent where $l$ and $b$ are in degrees.

This line passes through the $1\sigma$ error box of the peak position of 
HESS J1023-575, which is shifted from the center of the cluster by $\sim6$ 
arcmin. The arc is apparently symmetric with respect to the jet axis and 
is well centered on HESS J1023-575. It appears somewhat rim-brightened 
along the outer boundary and shows a marked crescent shape, suggesting 
part of a swept-up shell. For convenience we define
an axis $S$, which runs along the direction of the jet and passes through 
the peak position of HESS J1023-0575, and an axis $T$ which is perpendicular 
to $S$. Here the origin of both $T$ and $S$ is $(l, b) = (284^{\circ}.19, -0^{\circ}.39)$.

The velocity distributions of the two features are shown in two overlaid 
velocity channel maps (Figure 2a) and in a position-velocity map along $S$ (Figure 2b).

The jet is generally quiescent, having a velocity width of a few km s$^{-1}$ 
between $S \sim 0^{\circ}.1 - 0^{\circ}.6$ but show fairly large widths of 10 
km s$^{-1}$ at $S \sim 0^{\circ}.5$ and $0^{\circ}.7$, if all the features 
are assumed to be physically
associated. We tentatively divide the molecular gas towards the jet into the 
following four components;

J1: A narrow component at $S \sim 0^{\circ}.0 - 0^{\circ}.6$ located at 
$\sim29$ km s$^{-1}$ with a linewidth of 1 $ - $ 2 km s$^{-1}$, with little velocity gradient. 
This component has three local peaks at $S \sim 0^{\circ}.2, 0^{\circ}.3$, and $0^{\circ}.5$ 
with a winding shape towards $S \sim 0^{\circ}.1-0^{\circ}.3$.

J2: A broad component at $S \sim 0^{\circ}.6-1^{\circ}.0$ located over a velocity 
range of 19 $ - $ 30 km s$^{-1}$.

J3: A fairly broad and localized feature at $S \sim 0^{\circ}.5$ and $V_{\rm LSR} 
\sim 21 - 24$ km s$^{-1}$, towards one of the peaks of J1.

J4: The western component partially overlapping with the arc at 
$S \sim -0^{\circ}.2$ and $V_{\rm LSR} \sim 26$ km s$^{-1}$.

Although it is not conclusive, we suggest that the four components of the jet 
are physically related as discussed later. The $^{12}$CO linewidth of the arc 
is $\sim4$ km s$^{-1}$ centered at 26 km s$^{-1}$, and we infer that the possible 
expansion velocity of the arc is not large.

We shall here assume that the jet and arc are at $\sim5.4$ kpc with a 30\% error 
limit, and estimate the mass of each as follows, where an X factor of 
$2.0 \times 10^{20}$ cm$^{-2}$ (K km s$^{-1}$) is used to convert the $^{12}$CO 
intensity to molecular mass \citep{ber93}:  $M({\rm J1}) \sim 7.9 \times 10^3 M_{\solar}$, $M({\rm J2}) 
\sim 2.2 \times 10^4 M_{\solar}$, $M({\rm J3}) \sim 3.0 \times 10^3 M_{\solar}$, $M({\rm J4}) 
\sim 1.0 \times 10^3 M_{\solar}$, and $M({\rm arc}) \sim 2.0 \times 10^4 M_{\solar}$, 
respectively. The accuracy in mass is limited to $\pm 60\%$ because of the uncertainty in distance.

 We show the HI distribution at $V_{\rm LSR} = 25.56$ km s$^{-1}$ with a width of 0.82 
km s$^{-1}$ in Figure 3 \citep{mcc05}. The HI shows a clear sign of a hole towards Wd2 
and HESS J1023-575. Hereafter we shall call the HI feature surrounding the hole the HI shell. 
This shell is seen over a velocity range of 18 $ - $ 32 km s$^{-1}$. We note that the velocity 
width of the hole is rather small, a few km s$^{-1}$, suggesting that the possible expansion 
velocity is small at present. The HI hole has an elliptical shape with dimensions of 
$\sim29$ pc by $\sim14$ pc, elongated along the direction of S. It is noteworthy that the 
HI shell exhibits an intensity depression in its northern part, coincident with the molecular 
arc, suggesting that the arc and shell are physically related. This HI intensity depression 
is likely due to the conversion of HI into H$_2$. The jet is coincident with an elongated spur 
of HI extended to the east over $\sim1^{\circ}$, from $l \sim284^{\circ}.3$ to $285^{\circ}.0$, 
while on the west there is little HI beyond the arc. The HI mass of the cloud shown in Figure 3 
amounts to $\sim1.5 \times 10^5 M_{\solar}$ in the velocity range 18 $ - $ 32 km s$^{-1}$ for 
a conversion factor of $1.8 \times 10^{18}$ cm$^{-2}$ (K km s$^{-1})^{-1}$ \citep{dic90}.

\begin{figure}
  \begin{center}
    \FigureFile(70mm,70mm){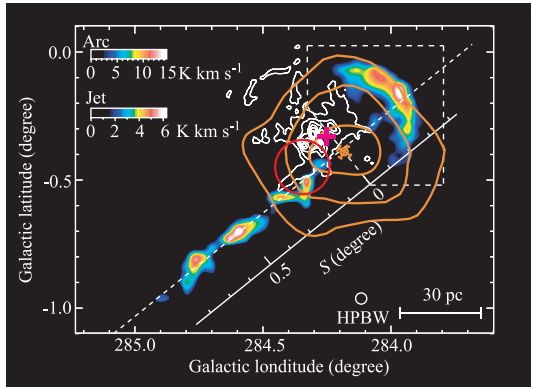}
  \end{center}
  \caption{The distribution of the $^{12}$CO emission integrated over a velocity range 
from 24 $ - $ 28 km s$^{-1}$ (the arc in the dashed region) and from 28 $ - $ 30 km s$^{-1}$ 
(the jet). The orange contours correspond to the TeV gamma ray source, HESS J1023-575 
\citep{aha07}. The red circle indicates  the location of the gravity centre of the low 
energy (MeV/GeV) gamma-ray source reported by the $Fermi$ collaboration \citep{abd09}. 
The white contours are the radio continuum \citep{whi97}. The cross indicates the position 
of Wd2. The axis S is defined along the direction of the jet with the position of HESS J1023-575.}\label{fig1}
\end{figure}

\begin{figure}
  \begin{center}
    \FigureFile(60mm,0mm){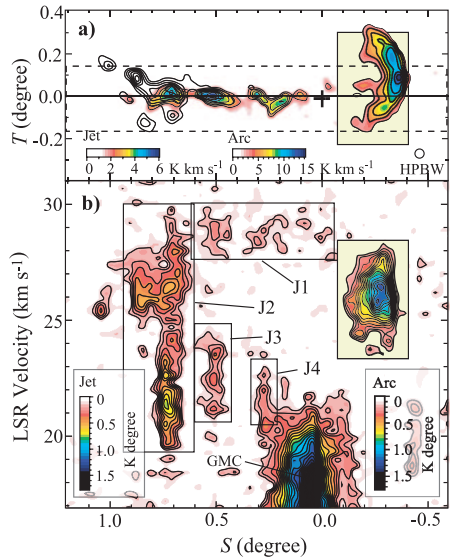}
  \end{center}
  \caption{Velocity distribution of the arc and the jet along the
axis of the jet. (a) Integrated intensity map in $^{12}$CO emission.
The velocity ranges are 24 $ - $ 28 km s$^{-1}$ for the arc and 28 $ - $ 30 km s$^{-1}$ for the jet.
Dashed contours show emission in the range 25 $ - $ 28 km s$^{-1}$ and are every 1.6 K
km s$^{-1}$ from 2.4 K km s$^{-1}$ ($\sim 3\sigma$). The blue mark indicates the position of
the TeV gamma ray source. (b) Position-velocity diagram taken along $S$
in figure 1. The horizontal axis indicates the coordinate of $S$ with the
position of the TeV gamma ray source as the origin. The definitions of
the components J1, J2 and J3 are described in the text.}\label{fig2}
\end{figure}

\section{Discussion}

The CO arc and HI shell suggest a spherical shock either due to a SN or stellar 
winds of WR stars. The secure lower limit for the kinetic energy involved in the 
arc is estimated to be $\sim2.0 \times 10^{47}$ ergs for a molecular mass of 
$\sim2.0 \times 10^4 M_{\solar}$ and an expansion velocity of 1 km s$^{-1}$. 
A more likely estimate is perhaps $\sim1 \times 10^{48}$ ergs or more if we 
add the expansion energy of the HI gas to the expansion energy of the gas 
traced in CO. A SN with mechanical energy $\sim10^{51}$ ergs is more than 
sufficient to account for the energetics of the observed molecular and atomic 
gas, even when taking into account the conversion efficiency into kinetic 
energy of $\sim0.1$ \citep{tho98}. There are many O stars in Wd2, whose 
collective mechanical wind energy may be large. For example 10 O-stars each 
with stellar-wind kinetic energy luminosities of $10^{36}$ erg s$^{-1}$  
yield $\sim10^{37}$ erg s$^{-1}$. Over a timescale of $\sim10^5$ years 
the total kinetic power could reach $\sim3 \times 10^{49}$ ergs. With 
the same conversion efficiency, 0.1, a kinetic energy of $\sim3 \times 
10^{48}$ ergs is available.

The one-sided molecular jet can be explained in terms of in-situ conversion 
of HI into molecular gas by shock compression as a result of the jet 
propagating through the atomic medium over a projected length of $\sim100$ pc. 
This is consistent with the huge mass of the molecular jet $-$ several 
times $10^4 M_{\solar}$ $-$ and the considerable amount of HI gas 
($\sim5.2 \times 10^4 M_{\solar}$ ) seen towards it. A lower limit for 
the time scale of the molecular jet is estimated to be $\sim 10^{5-6}$ yrs 
from the timescale of shocked CO formation (\cite{koy00}; \cite{ber04}). 
There are presently two other known candidates for molecular jets driven 
by high-energy jets $-$ SS433 and MJG348.5 \citep{yam08}. The physical 
parameters of the two molecular jets are estimated as follows; 
full length $\sim500$ pc, molecular mass $\sim10^4 M_{\solar}$, and 
minimum kinetic energy $\sim10^{48}$ ergs. For the Wd2 clouds, we roughly 
estimate such kinetic energy to be $\sim2.0 \times 10^{48}$ ergs from the mass, 
$\sim2.2 \times 10^4 M_{\solar}$, and the velocity span, $\sim10$ km s$^{-1}$, 
of the dominant component J2, which is somewhat larger than the above two cases. 
We note that an increased velocity dispersion at the far tip of the jet, as seen 
in J2, is also observed in SS433 and MJG348.5. This is interpreted as a result 
of the stronger dynamical interaction towards the tip of the jet, where the 
deceleration of the high-energy jet becomes most significant \citep{yam08}.

We now consider the origin of the jet and arc. Presently, there are two scenarios 
in which highly collimated flows of sufficient energy and length may be formed: 
(i) a highly anisotropic supernova explosion; (ii) a high-energy accretion-powered 
jet from a compact object such as in a microquasar. Anisotropic energy output is 
thought to have occurred in several known SNe including Cas A and W49B 
(\cite{hwa04}; \cite{mic08}). The expected speed of the jets 
is $>1000$ km s$^{-1}$, suggesting a travel time of $\sim10^4$ yrs over $\sim100$ pc 
if the jet axis is nearly perpendicular to the line of sight. Theoretical 
calculations indicate that a strongly magnetized and rotating neutron star formed 
in a SNe results in highly collimated jet \citep{bur07}. Such explosions are 
able to release $>10^{52}$ ergs in energy (\cite{kom07}), and the 
compact remnant is expected to be observable as a magnetar. The frequency of known 
magnetars is low, with only 12 magnetars \citep{mer08} amongst the 1500 or so 
known pulsars \citep{man04}, while jet-like features in historical SNRs such as 
Cas A suggest that the anisotropic SNe may not be very rare. To summarize, the 
anisotropic SNe scenario may well explain the jet and arc as caused by a single 
collimated SNe.

An alternative is that a conventional, isotropic SNe occurred in a binary system, 
leading to the formation of an accretion-powered jet.  For example, the microquasar 
SS433, exhibits an X-ray jet of 150 pc and molecular jet of 400 pc in length 
(\cite{kot98}; \cite{yam08}). In Wd2 the arc and shell might be formed by the 
SN explosion, and the molecular jet by a long-lived microquasar jet. The SS433 
jet has a momentum flow rate of $1.5 \times 10^{-7} M_{\solar}$ yr$^{-1}$ and 
can supply $10^{51}$ ergs in kinetic energy in $10^5$ yrs (\cite{kot98}; \cite{mar02}), 
sufficiently large to supply the kinetic energy above.

Concerning the origin of the gamma-ray emission, the intriguing alignment of 
the gravity centers of both the $Fermi$ and HESS source positions with the 
molecular jet and arc suggests a common (initial) event, e.g., a supernova explosion.  
The link could be direct, via radiation of relic particles produced at the SN explosion, 
or via on-going acceleration of particles by a relativistic object such as a remnant of 
the explosion, e.g.  supernova remnant shell, a pulsar wind nebula (PWN) or a microquasar jet.   
The first scenario favors a hadronic origin for the gamma-ray emission. Because of 
severe radiative (synchrotron and inverse Compton) losses, the relic ultrarelativistic 
electrons hardly could survive to the present epoch. Indeed, for any reasonable 
parameters characterizing the ambient medium, the lifetime of multi-TeV electrons 
is significantly shorter than the age of the system.  On the other hand, inverse 
Compton radiation of gamma-rays by continuously accelerated electrons, e.g., 
by a PWN, is a viable option. The gamma-ray luminosity is about $10^{35}$ erg s$^{-1}$.  
If the main target for IC gamma-rays is the 2.7 K CMBR, this mechanism would require 
acceleration of TeV electrons at a rate of $10^{37}$ erg s$^{-1}$, or an order of 
magnitude less,  if the optical radiation of stars plays a dominant role. 
The latter case can be realized when acceleration  and radiation of electrons take 
place not far from the stars. The extended character of the gamma-ray emission 
excludes a single star origin for the observed TeV emission.  In this case the 
extension of the gamma-ray source is effectively determined by the volume occupied 
by stars, and consequently the size of the source is expected to be energy-independent.  
In contrast to this scenario, for a single PWN we expect an energy-dependent morphology 
caused by radiative energy losses which do not allow the highest energy electrons 
to propagate to large distances.  Such an effect is observed in the PWN HESS J1825-137 
\citep{aha06}. In the case of a hadronic scenario we expect just the opposite 
dependence. Since generally the higher energy protons propagate faster than low energy 
protons, we might expect an increase of the angular size of the source with energy \citep{aha04}.   
Another test to distinguish between the hadronic and electronic models is the ratio of 
GeV and TeV gamma-ray fluxes. In the case of the IC mechanism, we expect suppression of 
low energy gamma-rays and a rather low GeV/TeV ratio. It is expected that future higher 
quality  GeV and TeV  data  should allow us to conduct detailed a quantitative study of 
the morphological and spectral characteristics of the radiation in different energy bands, 
which  could lead to definite conclusions. Here we would prefer to limit the discussion 
by noting that the total energy required to explain the TeV gamma-ray data reported by 
the HESS collaboration for a source located at a distance of  5 kpc is about 
$10^{50} (n/1 $cm$^{-3})^{-1}$ erg. The CO observations indicate a rather low density of 
gas in the region that coincides with the location of the TeV gamma-ray source (see above), 
most likely, $n<1$ cm$^{-3}$.   Then the total budget in relativistic protons is required 
as large as $10^{50}$ erg. This can be considered as an argument in favor of a
very strong SN explosion with total energy exceeding $10^{51}$ erg.

In summary, our observations conducted with the NANTEN telescope in the $^{12}$CO($J=1-0$) 
mm-wave emission led to the discovery of a spectacular jet and arc of molecular gas toward 
the young star cluster Westerlund 2 which may be the result of a powerful supernova explosion. 
Another consequence of this explosion could be the GeV and TeV gamma-ray sources located 
in the same region as reported recently by the $Fermi$ and HESS collaborations. Further 
detailed studies of spectral and morphological features of gamma-ray emission are 
requited to explore the links between these two phenomena.

\vspace{0.5cm}
The original NANTEN telescope was operated based on a mutual agreement between Nagoya 
University and the Carnegie Institution of Washington. This work is financially supported 
in part by a Grant-in-Aid for Scientific Research (KAKENHI) from the MEXT and from JSPS, 
in part, through the core-to-core program and by the donation from individuals and private companies.

\begin{figure}
  \begin{center}
    \FigureFile(70mm,70mm){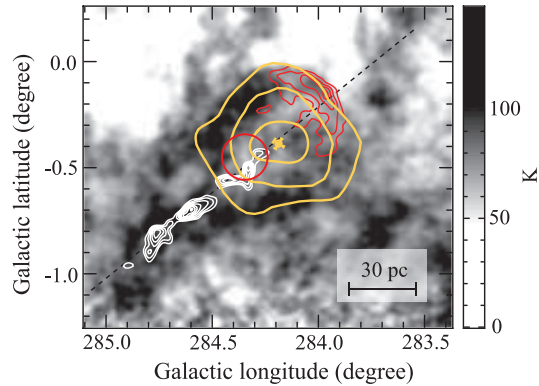}
  \end{center}
  \caption{Intensities of the HI and $^{12}$CO emission at a velocity of 25.56 km s$^{-1}$ 
\citep{mcc05}. The gray scale is HI emission with an intensity range of 0-150 K. Red and white 
contours show the integrated intensity of the arc and the jet. The yellow contours correspond 
to the TeV gamma ray source. The red circle shows the gravity center of the $Fermi$ source.}\label{fig3}
\end{figure}







\end{document}